**PAPER • OPEN ACCESS**

# Analytic auto-differentiable ΛCDM cosmography

View the article online for updates and enhancements.

# Analytic auto-differentiable ΛCDM cosmography

**Konstantin Karchev**

Theoretical and Scientific Data Science group,
Scuola Internazionale Superiore di Studi Avanzati (SISSA),
via Bonomea 265, 34136 Trieste, Italy

E-mail: kkarchev@sissa.it

**Abstract.** I present general analytic expressions for distance calculations (comoving distance, time coordinate, and absorption distance) in the standard ΛCDM cosmology, allowing for the presence of radiation and for non-zero curvature. The solutions utilise the symmetric Carlson basis of elliptic integrals, which can be evaluated with fast numerical algorithms that allow trivial parallelisation on GPUs and automatic differentiation without the need for additional special functions. I introduce a `PyTorch`-based implementation in the `phytorch.cosmology` package and briefly examine its accuracy and speed in comparison with numerical integration and other known expressions (for special cases). Finally, I demonstrate an application to high-dimensional Bayesian analysis that utilises automatic differentiation through the distance calculations to efficiently derive posteriors for cosmological parameters from up to $10^6$ mock type Ia supernovæ using variational inference.

**Keywords:** Machine learning , supernova type Ia - standard candles

**ArXiv ePrint:** 2212.01937

https://doi.org/10.1088/1475-7516/2023/07/065

## Contents

## 1 Introduction

The deep-learning revolution brought about by automatic differentiation and general-purpose parallel computing on graphics processing units (GPUs) has motivated the development of a number of new high-performance automatically differentiable simulators (and emulators) across cosmology: e.g. for large-scale structure [31, 32, 38, 46, 47, 56, 57], weak lensing [4], strong lensing [17, 24, 29, 41], gravitational waves [18], and in related fields [1, 26, 30, 36, 43, 48, 51, 61, 73, 76, 80]. Having access to gradients through the simulator then enables general high-dimensional likelihood-based analyses with Hamiltonian Monte Carlo (HMC) [21, 34] or variational inference (VI) [35, 40, 66] and can be used in the context of likelihood-free simulation-based inference to speed up the training of neural networks through an additional loss term [5, 15, 79].

All of the aforementioned cosmological simulators require calculating cosmographic distances, which are also a key ingredient in the modelling and data analysis of standard candles, sirens, and rulers, volumetric rates and densities, the cosmic microwave background radiation, Ly $\alpha$ forests in quasar spectra, as well as in the studies of galaxy properties and evolution. In the general case, cosmographic calculations require evaluating integrals numerically, which is both slow and not trivially parallelisable, while requiring a further numerical integration for the gradient calculation.[1]

---

[1]See e.g. the `torchdiffeq` package [16] and the background routines of `jaxcosmo` (https://jax-cosmo.readthedocs.io/en/latest/_modules/jax_cosmo/background.html#radial_comoving_distance).

Here I describe a unified analytic approach to distance calculations in a general ΛCDM+radiation cosmology, formulated in terms of the Carlson elliptic integrals. These special functions were introduced by [7] as an alternative to Legendre's solutions to elliptic integrals. They are more suited to analytical rather than geometrical problems and offer two further advantages: first, fast and precise numerical algorithms for their evaluation exist [10], which makes their implementation for GPUs straightforward; and second, their derivatives can also be evaluated analytically without the use of additional special functions. Furthermore, a general reduction scheme [11] allows all cosmographic distances to be calculated within the same framework.

This paper is structured as follows. Section 2 describes cosmographic distances in the ΛCDM+radiation setting and discusses the Carlson elliptic integrals and their derivatives. The explicit solutions are presented in section 3, with simplified expressions for the radiationless case given in appendix B.1. A GPU-enabled automatically differentiable implementation in the `phytorch.cosmology` package is introduced in section 3.2, and its accuracy and speed are briefly examined. Finally, an application to gradient-based cosmological inference from standard candles is demonstrated in section 4, and conclusions are drawn in section 5.

## 2 Background

### 2.1 ΛCDM cosmology

In the homogeneous and isotropic cosmological model of Friedmann, Lemaître, Robertson, and Walker (FLRW), the evolution of a universe is described by the dimensionless Hubble parameter $E(t)$, whose form is determined by the Friedmann equations based on the universe's composition.

From a few seconds after the Big Bang onward, the large-scale evolution of the Universe is well described by a homogeneous non-interacting mixture of *radiation*,[2] *matter*, and *dark energy* in the form of a cosmological constant (Λ), constituting the ΛCDM+radiation cosmological model. The amounts present of each component (including the effective contribution of spatial *curvature*) are quantified by the dimensionless *density parameters* $\{\Omega_i(z)\}_{i\in\{r,m,\Lambda,k\}}$, which have different evolution laws:

$$\Omega_{\rm r}(z) \propto \Omega_{\rm r0}(1+z)^4, \quad \Omega_{\rm m}(z) \propto \Omega_{\rm m0}(1+z)^3, \quad \Omega_{\rm k}(z) \propto \Omega_{\rm k0}(1+z)^2, \quad \Omega_{\Lambda}(z) \propto \Omega_{\Lambda 0}, \quad (2.1)$$

with the time axis parametrised by the *cosmological redshift*[3] $z$ and $\Omega_{i0}$ the dimensionless density parameter of component $i$ at the present time ($z = 0$).[4] The parameters are defined to sum to one at all times, which is often used to express the curvature parameter as $\Omega_{\rm k} = 1 - \sum_{i\neq k} \Omega_i$. Measured values for the current epoch are $\Omega_{\rm m0}$ and $\Omega_{\Lambda 0}$ of order unity with a dominance of dark energy, $\Omega_{\rm r0} \approx 10^{-5}$, and $\Omega_{\rm k0} \approx 0$ consistent with a spatially flat universe [60]. By the first Friedmann equation ([59], eq. (13.3)),

$$E^2(z) = \Omega_{\rm r0}(1+z)^4 + \Omega_{\rm m0}(1+z)^3 + \Omega_{\rm k0}(1+z)^2 + \Omega_{\Lambda 0}. \quad (2.2)$$

---

[2]More formally, a relativistic component, which includes all sufficiently light (or massless) particles that have relativistic energies throughout the whole history of the universe, e.g. sufficiently light neutrinos.

[3]The re-parametrisation time ↔ redshift is only valid if the scale factor $a(t)$ is monotonic. Otherwise, one can split the evolution in phases and still apply the transformation piecewise, but we will not concern ourselves with such "big bounce" scenarios.

[4]In general, a component is described by its, possibly evolving, *equation of state* $\mathfrak{w}(z)$, which leads to an evolution law $\propto \exp\left[3 \int (1 + \mathfrak{w}(z))\,\mathrm{d}z/(1+z)\right]$. Our method for calculating distances requires $E^2$ to be a polynomial of degree at most 4, which excludes evolving dark energy models as well as a general equation of state, restricting it to $\mathfrak{w} \in \{-1, -2/3, -1/3, 0, 1/3\}$.

– 2 –

### 2.1.1 Cosmography

Three fundamental distance integrals: the *radial comoving distance*, *time coordinate*, and *absorption distance*, are defined, respectively, as

$$\chi(z_1, z_2) = \frac{c}{H_0} \int_{z_1}^{z_2} \frac{\mathrm{d}z}{E(z)}, \tag{2.3}$$

$$t(z_1, z_2) = \frac{1}{H_0} \int_{z_1}^{z_2} \frac{\mathrm{d}z}{(1+z)E(z)}, \tag{2.4}$$

$$d_{\mathrm{abs}}(z_1, z_2) = \frac{c}{H_0} \int_{z_1}^{z_2} \frac{(1+z)^2}{E(z)} \mathrm{d}z, \tag{2.5}$$

(ref. [59], eqs. (13.40), (13.9), and (13.42)), where $H_0$ is the current value of the Hubble parameter, and $c$ is the speed of light. From them physical observables can be derived easily: the comoving distance (modified geometrically to take into account curvature) is related via the scale factor to the *angular diameter/size* and *luminosity* "distances" ([59], eqs. (13.47) and (13.57)) and to the *comoving volume* ([59], eq. (13.60)) used in standard candle/siren/ruler studies; setting appropriate limits in eq. (2.4) gives the *lookback time* ($\int_0^z \mathrm{d}t$) or an object's *age* ($\int_z^\infty \mathrm{d}t$); and finally, $d_{\mathrm{abs}}$ is related to the *intersection probability* of the line of sight with objects of constant comoving number density and proper cross section (used in modelling the Ly $\alpha$ forest).

Analytical solutions to eqs. (2.3) to (2.5) in terms of *special functions*, which can be evaluated to arbitrary precision by numerical algorithms that do not (directly) reference the defining integral, have been discussed in the literature only for certain special cosmologies. For example, the comoving distance in a flat $\Lambda$CDM universe can be expressed using the Gauss hypergeometric function [2], the Legendre elliptic integrals [22, 55, 77, 78], which are also applicable in the non-flat case ([23, 69], see also references therein), and the Carlson elliptic form [49].[5] Ref. [20] presented a solution valid also in the presence of radiation and curvature that makes use of the Weierstrass elliptic function,[6] and finally, ([70], appendix B) used Carlson's basis to solve for the time coordinate.[7]

The formulæ presented in this work are the first to unify the different distance calculations in one framework applicable in the most general case of a $\Lambda$CDM cosmology with non-zero curvature and in the presence of radiation. They also have the advantages of fast and parallelisable numerical evaluation and easy to compute gradients with respect to the cosmological parameters.

## 2.2 The Carlson symmetric elliptic integrals

The integrals eqs. (2.3) to (2.5) cannot be expressed for general $\{\Omega_{i0}\}$ as elementary functions. However, if $E^2(z)$ is a polynomial of degree up to four, they are instances of *elliptic integrals*, which can be reduced to a linear combination of a small set of *basis integrals*, e.g. Legendre's elliptic integrals [6, 45]. The resulting expressions are present in most comprehensive tables of integrals (e.g. [6, 27]).[8]

---

[5]Liu et al. [49] mentioned the applicability of Carlson's formulation to the non-flat case but did not elaborate.

[6]Which is, in my opinion, more of theoretical than of practical importance since methods for its numerical evaluation are hard to come by.

[7]Valkenburg [70] approached the problem from the perspective of the slightly more general Lemaître-Tolman-Bondi (LTB) metric, describing an evolving spherically symmetric isolated collection of "dust".

While Legendre's formulation is well suited to geometric problems, an alternative basis, the Carlson symmetric form ([7], see also [12], section 19), is the natural choice when dealing with rational functions. By preserving the original permutation symmetry in the polynomial roots, it unifies the different cases that select the correct branches of the square roots involved. Furthermore, for the Carlson basis integrals there are efficient numerical algorithms with guaranteed fast convergence to an arbitrary precision in the whole complex plane[9] [10]. In fact, these algorithms are the basis for some numerical implementations of Legendre's integrals ([62], section 6.11).

The Carlson integrals of the first and third kind, which form the basis for reduction, are defined by:

$$R_F(x_1, x_2, x_3) \equiv \frac{1}{2} \int_0^\infty \frac{\mathrm{d}z}{\sqrt{(z+x_1)(z+x_2)(z+x_3)}}, \tag{2.6}$$

$$R_J(x_1, x_2, x_3, w) \equiv \frac{3}{2} \int_0^\infty \frac{\mathrm{d}z}{(z+w)\sqrt{(z+x_1)(z+x_2)(z+x_3)}}. \tag{2.7}$$

It is useful to define also the degenerate versions:

$$R_C(x_1, x_2) \equiv R_F(x_1, x_2, x_2), \tag{2.8}$$

$$R_D(x_1, x_2, x_3) \equiv R_J(x_1, x_2, x_3, x_3). \tag{2.9}$$

The functions are well-defined for all complex $x_1, x_2, x_3$ except the non-positive reals (for which the integrand has poles along the integration path) and for all non-zero $w$ (the Cauchy principal value is assumed if $w \in \mathbb{R}_{<0}$). $R_F$ and $R_J$ are symmetric in $\{x_1, x_2, x_3\}$, while $R_D$ is only symmetric in $\{x_1, x_2\}$, and $R_C$ is not symmetric.

Computing derivatives of the Carlson integrals is closed, i.e. does not require any other special functions:

$$\frac{\partial R_F}{\partial x_3} = -R_D/6, \tag{2.10}$$

(ref. [12], chapter 19, eq. (19.18.1)), and

$$\frac{\partial R_J}{\partial x_3} = \frac{1}{2} \frac{R_J - R_D}{w - x_3}, \tag{2.11}$$

$$\frac{\partial R_J}{\partial w} = \frac{3}{2} \left\{ \frac{w^2 R_F - 2w R_D + \prod_{i=1}^3 \sqrt{x_i}}{w \prod_{i=1}^3 (w - x_i)} - \left( \sum_{i=1}^n \frac{1}{w - x_i} \right) \frac{R_J}{3} \right\}, \tag{2.12}$$

ref. [75], where $R_F$ and $R_D$ are evaluated at $(x_1, x_2, x_3)$, and $R_J$ at $(x_1, x_2, x_3, w)$. Derivatives with respect to $x_1$ and $x_2$ are obtained by symmetry, while those of $R_C$ and $R_D$ via their respective definitions and the chain rule. In eqs. (2.11) and (2.12) the limits have to be explicitly implemented when arguments are repeated (e.g. $w = x_i$ or $x_i = x_j$: see [75] for details).[10]

---

[8]For reference, in [27] the comoving distance integral is eq. (3.147) (or eq. (3.131) in the radiation-less case), and the time integral is eq. (3.151) (or eq. (3.137)). The absorption distance integral can be reduced to a combination of the latter two and eq. (3.148) (or eq. (3.132)) via eq. (250.01) (or eq. (230.01)) of [6].

[9]Except for certain cases of $R_J$, when the algorithm is not *guaranteed* although often correct in practice.

[10]The same applies to higher-order derivatives. In general, the limits of repeated arguments either have to be evaluated numerically or hard-coded.

Reductions of particular elliptic integrals to the Carlson form were tabulated by [8, 9], while general reduction schemes, which can be easily implemented in computer algebra systems or purely numerically, were later presented by [11] and [28]. Since the Carlson basis involves integrals with fixed bounds, the associated reduction scheme applies to definite integrals, while reduction to Legendre's form usually considers the indefinite version, thus requiring two evaluations of the resulting expression or additional manipulation.

## 3 A unified analytic approach to distance calculations

In Carlson's tables [8] elliptic integrals are standardised as

$$\mathcal{I}(\mathbf{p}, \mathbf{a}, z_1, z_2) \equiv \int_{z_1}^{z_2} \prod_{i=1}^{n} (z + a_i)^{p_i/2} \, \mathrm{d}z \,, \tag{3.1}$$

parametrized by the $n$-tuples (i.e. ordered sets) $\mathbf{p} \equiv [p_i]_{i=1}^n$ and $\mathbf{a} \equiv [a_i]_{i=1}^n$. The $p_i$ must be integers, and exactly $m = 3$ or $m = 4$ of them must be odd (corresponding to $E^2$ being a degree-3 or -4 polynomial). Each of the factors $z + a_i$ is assumed to be positive on the interval $(z_1; z_2)$.[11]

To make use of the Carlson basis, one must, therefore, factorise $E^2(z)$. Since in the $\Lambda$CDM+radiation framework described above it is a degree-four polynomial (cf. eq. (2.2)), this can be achieved using well-known explicit algebraic formulæ ([64], section 1.11(iii)) or established numerical methods, e.g. via the companion matrix[12] ([62], section 9.5). We write the resulting factorisation as

$$E^2(z) = \alpha_m \prod_{j=1}^{m} (z - r_j), \tag{3.2}$$

where $m = 4$, $\alpha_m = \Omega_{\mathrm{r}0}$ is the leading coefficient of the polynomial,[13] and $[r_j]_{j=1}^m$ are the, generally complex, roots of $E^2(z) = 0$. Since this effectively re-parametrises the problem with $\{\Omega_i\} \to \{r_i\}$, one needs to also compute the gradient of the root calculation, which is discussed in appendix A.

Using eq. (3.2), the integrands of eqs. (2.3) to (2.5) can be recast in the form

$$\alpha_m^{-\frac{1}{2}}(z+1)^{\frac{p_{m+1}}{2}} \left[ \prod_{j=1}^{m} (z - r_j) \right]^{-\frac{1}{2}} \quad \to \quad \alpha_m^{-\frac{1}{2}}(z+1)^{\frac{p_{m+1}}{2}} \prod_{j=1}^{m} (z - r_j)^{-\frac{1}{2}}, \tag{3.3}$$

with $p_{m+1} \in \{0, -2, 4\}$ respectively. The factorisation of the square root into a product of square roots is valid since $E^2$ is real.

For physical solutions, we require that no root lie on the integration path, so that the integrand does not diverge, i.e. that $z = 0$ is smoothly connected to the Big Bang at $z \to \infty$. Parameter values for which this is not satisfied, exclusively with $\Omega_{\mathrm{m}0} < 0$ or $\Omega_{\Lambda 0} > 1$, are depicted in red in figure 1.

---

[11]In fact, the tables consider linear factors $b_i z + a_i$, but we will always have $b_i = 1$. Since one can always re-define $a_i \to a_i/b_i$ and pre-multiply the whole expression appropriately, having general $b_i$ accounts only for some necessary sign changes in the formulæ when a factor is negative on the whole interval. To simplify notation, therefore, we have explicitly set $b_i = 1$ in all formulæ.

[12]Matrix formulations are advantageous since machine learning libraries and hardware are usually highly optimised for the operations involved. In some cases, numerical stability might even be better than when using the analytic formulæ directly.

[13]One can consider equivalently the radiation-less case, for which $m = 3$ and $\alpha_m = \Omega_{\mathrm{m}0}$.

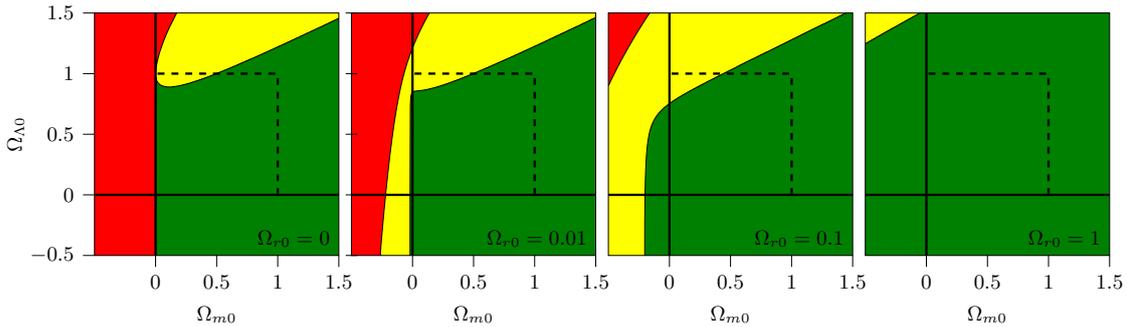

**Figure 1.** Nature of the roots of $E^2(z) = 0$ for different values of the leading coefficient $\Omega_{r0}$: when all roots have negative real part (green region), distance calculations are numerically stable; if there is a complex-conjugate pair with positive real part (yellow region), the numerical stability of evaluating eqs. (3.8) and (3.9) is affected by the choice of permutation of the roots; finally, the red region comprises the parameters of unphysical universes without a Big Bang, for which $E^2$ has two positive real roots, and so the Hubble parameter diverges at finite redshift.

### 3.1 Explicit formulæ in terms of Carlson's elliptic integrals

Comparing eq. (3.3) with eq. (3.1), we can identify $a_j = -r_j$ for $1 \leq j \leq m$ and $a_{m+1} = 1$. Using $\cdot \| \cdot$ to denote concatenation of tuples, we can write this as $\mathbf{a} = (-\mathbf{r}) \| [1]$. Equations (2.3) to (2.5) then become

$$\chi(z_1, z_2) = \frac{c}{H_0} \alpha_m^{-\frac{1}{2}} \, \mathcal{I}(-\mathbf{1}_m, -\mathbf{r}, z_1, z_2), \tag{3.4}$$

$$t(z_1, z_2) = \frac{1}{H_0} \alpha_m^{-\frac{1}{2}} \, \mathcal{I}((-\mathbf{1}_m) \| [-2], (-\mathbf{r}) \| [1], z_1, z_2), \tag{3.5}$$

$$d_{\text{abs}}(z_1, z_2) = \frac{c}{H_0} \alpha_m^{-\frac{1}{2}} \, \mathcal{I}((-\mathbf{1}_m) \| [4], (-\mathbf{r}) \| [1], z_1, z_2), \tag{3.6}$$

where $-\mathbf{1}_m \equiv [-1]_{i=1}^m$ is an $m$-tuple of negative ones.

The final formulæ for the comoving distance, time coordinate, and absorption distance in terms of the Carlson symmetric integrals (from [8], eqs. (3.1), (2.35), and (2.33)) are

$$\chi(z_1, z_2) = \frac{c}{H_0} \Omega_{r0}^{-\frac{1}{2}} \times 2\Delta z R_f(u_{12}^2, u_{13}^2, u_{23}^2), \tag{3.7}$$

$$t(z_1, z_2) = \frac{1}{H_0} \Omega_{r0}^{-\frac{1}{2}} \times \frac{2\Delta z}{r_i + 1} \times \tag{3.8}$$
$$\times \left\{ R_f(u_{12}^2, u_{13}^2, u_{23}^2) - \left[ \frac{(\Delta z)^2}{3} \frac{d_{ij} d_{ik} d_{il}}{d_{i5}} R_J(u_{12}^2, u_{13}^2, u_{23}^2, u_{i5}^2) + R_C(s_{i5}^2, q_{i5}^2) \right] \right\},$$

$$d_{\text{abs}}(z_1, z_2) = \frac{c}{H_0} \Omega_{r0}^{-\frac{1}{2}} \times \Delta z \times \left\{ \frac{1}{\Delta z} \left[ \frac{\sqrt{z - r_i}\sqrt{z - r_j}\sqrt{z - r_k}}{\sqrt{z - r_l}} \right]_{z_1}^{z_2} \right. \tag{3.9}$$
$$+ \left[ 2(r_i + 1)^2 - d_{ij} d_{ik} \right] R_f(u_{12}^2, u_{13}^2, u_{23}^2)$$
$$+ d_{jl} d_{kl} \left[ \frac{(\Delta z)^2}{3} d_{ij} d_{ik} R_d(u_{ij}^2, u_{ik}^2, u_{il}^2) + \frac{\sqrt{z_2 - r_i}\sqrt{z_1 - r_i}}{\sqrt{z_2 - r_l}\sqrt{z_1 - r_l}} \frac{1}{u_{il}} \right]$$
$$\left. + \left( \sum_{\mu=1}^m r_\mu + m \right) \left[ \frac{(\Delta z)^2}{3} \frac{d_{ij} d_{ik} d_{il}}{d_{i0}} R_J(u_{12}^2, u_{13}^2, u_{23}^2, u_{i0}^2) + R_C(s_{i0}^2, q_{i0}^2) \right] \right\},$$

where $\{i, j, k, l\}$ is any permutation of $\{1, 2, 3, 4\}$, $\Delta z \equiv z_2 - z_1$, $d_{ij} \equiv r_j - r_i$ (with the special cases $d_{i0} \equiv -1$ and $d_{i5} \equiv -(1 + r_i)$), and

$$u_{ij} \equiv \sqrt{z_2 - r_i}\sqrt{z_2 - r_j}\sqrt{z_1 - r_k}\sqrt{z_1 - r_l} + \sqrt{z_1 - r_i}\sqrt{z_1 - r_j}\sqrt{z_2 - r_k}\sqrt{z_2 - r_l},$$

$$u_{i5}^2 \equiv u_{ij}^2 - (\Delta z)^2 \times d_{ik} d_{il} \frac{d_{j5}}{d_{i5}} \quad \rightarrow \quad u_{i0}^2 \equiv u_{ij}^2 - (\Delta z)^2 \times d_{ik} d_{il},$$

$$s_{i5}^2 \equiv q_{i5}^2 + (\Delta z)^2 \times d_{k5} d_{l5} \frac{d_{j5}}{d_{i5}} \quad \rightarrow \quad s_{i0}^2 \equiv q_{i0}^2 + (\Delta z)^2,$$

$$q_{i5}^2 \equiv \frac{(z_1 + 1)(z_2 + 1)}{(z_1 - r_i)(z_2 - r_i)} u_{i5}^2 \quad \rightarrow \quad q_{i0}^2 \equiv \frac{u_{i0}^2}{(z_1 - r_i)(z_2 - r_i)}.$$

(The quantities $u_{i0}$, $s_{i0}$, $q_{i0}$ are calculated as $u_{i5}$, $s_{i5}$, $q_{i5}$ but with all factors like $\square + 1 \to 1$, which corresponds to treating $(z + a_5)$ as identically one instead of $(1 + z)$. Note also that I have modified slightly the expressions from [8] by taking factors of $\Delta z$ outside the function arguments where possible via the homogeneity relations ([8], eq. (1.4)).)

Notice that eqs. (3.8) and (3.9) are not explicitly symmetric in $\{i, j, k, l\}$, i.e. in the ordering of the polynomial roots. Even though the result does not in principle depend on the exact chosen permutation, owing to the properties of the Carlson integrals, numerical evaluation might still be affected.[14] After some experimentation, I have found that simply picking $r_i$ and $r_j$ to be the biggest roots[15] by absolute value gives good results across parameter space and redshift range for all three distances and so defer an in-depth investigation to future work.

### 3.2 Implementation: `phytorch.cosmology`

The expressions eqs. (3.7) to (3.9) have been implemented in the `python` package `pytorch.cosmology`, a sub-library of `phytorch`,[16] which includes CUDA kernels for the Carlson elliptic integrals, interfaced through `PyTorch` [58], along with the respective gradient expressions (eqs. (2.10) to (2.12)) for use with `PyTorch`'s automatic-differentiation engine.[17] The Legendre integrals are also available through their relations to the Carlson form, which enables implementations of the other known cosmographic distance formulæ valid in special cosmologies. Finally, the general reduction schemes of [11] and [28] are implemented in `phytorch` for completeness.

#### 3.2.1 Accuracy

I briefly verify the `phytorch.cosmology` implementation against numerical integration[18] for a range of redshifts, considering two setups: a flat radiation-less universe and a non-flat one with a small amount of radiation present. As shown in figure 2, there is generally good agreement in calculating the comoving distance and lookback time using expressions from this paper (both in the Legendre and Carlson formulations).[19] Because of the sheer size of eq. (3.9), the absorption distance calculation is prone to small roundoff errors, which also strongly depend on the permutation of roots chosen. Lastly, the accuracy of the root calculation itself, which is not trivial when the coefficients span different orders of magnitude, affects the distance estimates, and in that respect numerical root-finding methods worked better than the algebraic formulæ.

---

[14]For example, for certain orderings, but not others, the arguments of $R_J$ might fall outside of the region in which Carlson's algorithm [10] is guaranteed. Or numerical instabilities might lead to spurious erroneous branch selection for e.g. $R_C$, especially when there are roots with positive real part (i.e. for cosmological parameters from the yellow region of figure 1).

[15]In fact, the biggest roots of $E^2(z+1)$, instead of $E^2(z)$, from which 1 is subtracted after ordering. This usually results in singling out in eqs. (3.8) and (3.9) the two negative real roots (if they are present).

[16]https://github.com/kosiokarchev/phytorch.

[17]Note that automatic differentiation differs from numerical differentiation in that the engine computes a gradient *analytically* using the chain rule and provided expressions for the derivatives of the individual components of complicated expressions, instead of using a finite difference method.

[18]Using the `scipy.integration.quad` function of `SciPy` [72], which itself wraps the adaptive Gauss-Kronrod quadrature routine of QUADPACK.

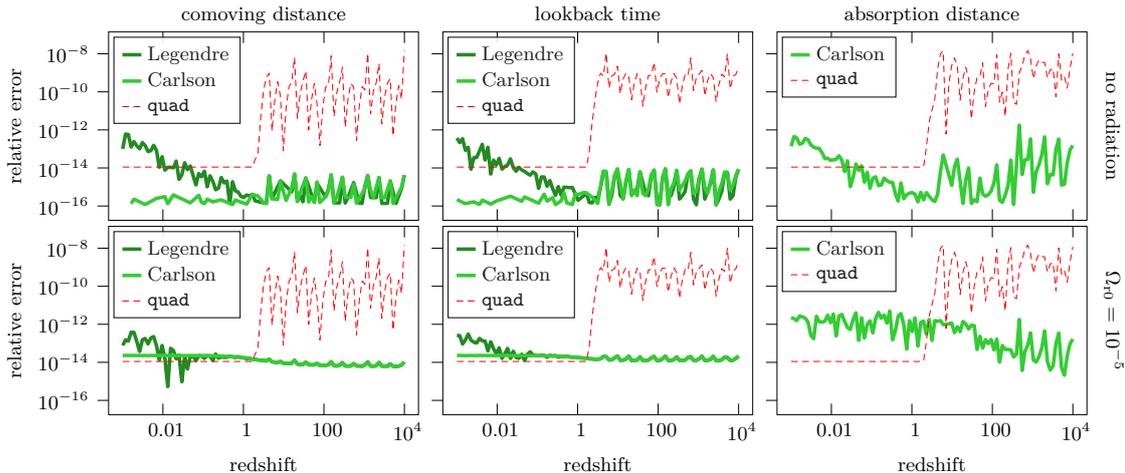

**Figure 2.** Relative difference between distance calculations (from $z_1 = 0$ to the redshift on the abscissa) using the formulæ from this paper and using numerical quadrature. For comparison, the dashed red lines show the reported error estimate of `quad` itself (divided by the actual value). The setup for the top row is a flat radiation-less universe with $\Omega_{m0} = 0.3$ and $\Omega_{\Lambda 0} = 0.7$, while for the bottom row an additional radiation component with $\Omega_{r0} = 10^{-5}$ is present. Expressions labelled "Carlson" are those in eqs. (B.1) to (B.3) (top row) and eqs. (3.7) to (3.9) (bottom row), while "Legendre" labels eqs. (B.4) and (B.5) (top row) and eqs. (B.6) and (B.7) (bottom row).

### 3.2.2 Speed

Although speed is not the primary focus of the current study, I examine the performance of the analytic formulæ and of numerical integration in the same two setups as in subsection 3.2.1. In the radiation-less, and more importantly, flat case, I also include the hypergeometric solution[20] [2].

All tests were performed in a single-threaded setting on a CPU. While analytic implementations are trivially parallelisable on multi-threaded CPUs and on GPUs, parallel numerical integration is easy for different limits ($z_1$, $z_2$) but harder for differing cosmological parameters since it requires multiple evaluations on the integration grid, and this is why it usually does not benefit much from GPU acceleration.

The results are summarised in figure 3. Carlson's formulation exhibits a step-like speed-up for small ranges of integration due to the decreasing number of iterations needed in the numerical algorithm. The bahaviour of Legendre's formulation is, instead, reversed, and the speed-up occurs for large upper limits. Numerical integration behaves similarly and, logically, is highly performant for small ranges when the integrand is approximately constant. For high redshifts, though, it can be more than an order of magnitude slower than analytic formulæ; however, this difference may well be due to the very different implementations used. Finally, as [2] observed, the hypergeometric solution has complicated behaviour with a drop in performance around $z = 1$ due to the nature of the numerical algorithm.[21]

---

[19]As seen from eqs. (B.4) to (B.7), the Legendre formulation requires taking the difference between the elliptical integrals evaluated at the two limits of integration, which is prone to roundoff errors when $z_1$ and $z_2$ are close. In the low-redshift limit, though, Taylor expansions (often linear) are anyway employed.

[20]A CUDA kernel for the Gauss hypergeometric function $_2F_1$ is also available in `phytorch`, albeit currently not automatically differentiable.

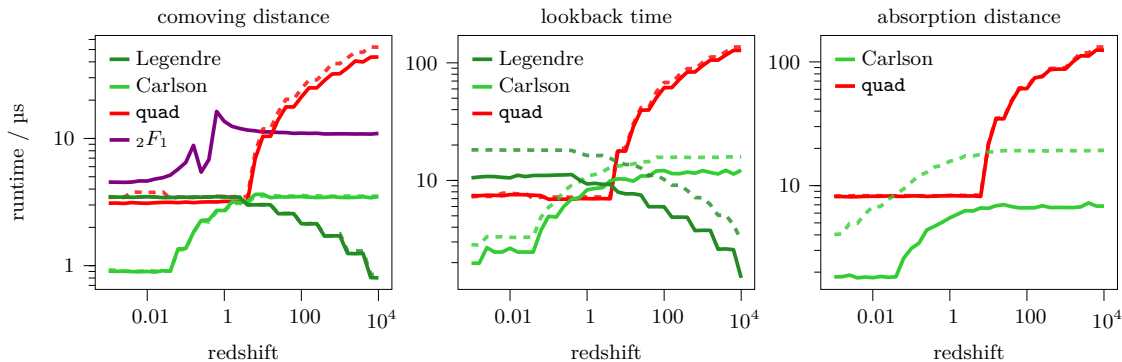

**Figure 3.** Speed comparison of different expressions / methods of calculating cosmographic distances (from $z_1 = 0$). The universe considered is flat with $\Omega_{m0} = 0.3$, $\Omega_{\Lambda 0} = 0.7$. (The dashed lines include, in addition, radiation with density $\Omega_{r0} = 10^{-5}$.) The expressions in eqs. (3.7) to (3.9) (and eqs. (B.1) to (B.3)) are labelled "Carlson", while "Legendre" labels their re-formulation in terms of Legendre integrals (eqs. (B.4) to (B.7), corresponding to the solution of [55]). "$_2F_1$" is equivalent to the formula of [2], and "quad" stands for numerical integration using SciPy. All experiments use 1 CPU thread.

The same conclusions apply also when the radiation term is included in $E^2(z)$ (see the dashed lines in figure 3) with almost no noticeable difference in the evaluation of the comoving distance, a factor $< 2$ slowdown for the time coordinate and less than an order of magnitude for the absorption distance. Since eqs. (3.8) and (3.9) are not symmetric in the roots of $E^2(z)$, even though the final numerical result is the same, using different permutations significantly affects the runtime since the required iterations to calculate $R_J$ are different.

## 4 Application: Bayesian inference of standard candles

Probably the most prominent application of cosmographic distances is in inference using standard(isable) candles (a "neoclassical test of cosmology" according to [59]). A standardisable candle is an object whose *intrinsic* brightness (absolute magnitude) can be derived from other observables. Its *observed* brightness, then, can be used to derive the luminosity distance to it via the inverse square law.[22] By obtaining in addition the cosmological redshift, one can in principle constrain the parameters of the cosmological model.

---

[21]Since phytorch.cosmology uses an alternative set of arguments to $_2F_1$, the behaviour at low and high $z$ is reversed with respect to [2]. Curiously, possibly also because of the slightly extended numerical algorithm capable of handling complex numbers, a slight recovery of performance is observed when the last argument $(\Omega_{m0}/\Omega_{\Lambda 0})(1+z)^3 \approx 1$. Finally, in this experiment the elliptic formulations significantly out-perform $_2F_1$, contrary to [2]'s result, again due to differences in the numerical implementation.

[22]In fact, as described below, analysis is usually performed with the distance modulus, which is defined as the difference between observed and absolute magnitude and is, therefore, related to the luminosity distance by $\mu \equiv m - M = 2.5 \log_{10}[(d_L/10\,\text{pc})^2]$.

Current analyses of type Ia supernovæ (SNæ Ia), the most widely used standardisable candle, rely on sophisticated Bayesian hierarchical models (BHMs) to infer, on one hand, global properties of the SN Ia population, and, on the other, intrinsic characteristics of individual objects [33, 52–54, 65, 68]. While applying these models to current datasets (e.g. [67]), containing $\sim 1000\,\text{SNæ Ia}$, is still feasible with sampling-based methods like Gibbs sampling and HMC,[23] the necessity to explore the whole parameter space, whose size scales with the number of observed SNæ, makes traditional inference techniques prohibitively expensive for analysis of the up to $10^6\,\text{SNæ Ia}$ expected in the near future from LSST [37, 44].

The burden of high dimensionality can be alleviated[24] by using modern gradient-based Bayesian techniques like variational inference (VI), in which the posterior is approximated by a tractable distribution, commonly referred to as the *variational proposal* or *guide*, implemented either in a simple analytic form, or with a neural network-based density estimator. Parameters of the proposal are determined by maximising a specifically designed function, called the evidence lower bound (ELBO), which includes the model likelihood conditioned on the data. Automatic differentiation through the model is, thus, the key to making VI a viable technique for high-dimensional inference since it enables optimisation with stochastic gradient descent. Even though one is still required to infer *all* model parameters jointly, it is possible to purposefully ignore some correlations and/or derive only point or Gaussian *a posteriori* estimates for certain parameters, thus simplifying the structure of the proposal and the learning task, if one is careful not to introduce biases (or is at least conscious of them). For a recent review of VI and the mathematical formalism, refer to [81]; see also ([3], subsection 11.3.2) for further discussion.

In this demonstration, I use a very simplified model for supernova cosmology, presuming to have measured the redshifts $\hat{z}^s$ and derived standardised distance moduli $\hat{\mu}^s$, of $N$ type Ia supernovæ.[26] For the latter I assume Gaussian likelihoods with equal and independent uncertainties: $\hat{\mu}^s | \mu^s \sim \mathcal{N}(\mu^s, \sigma_\mu^2)$, where $\sigma_\mu$ is the combined uncertainty due to residual scatter after standardisation and measurement noise. For the redshifts I adopt the toy model of [63], which also has a Gaussian likelihood albeit with a redshift-dependent variance: $\hat{z}^s | z^s \sim \mathcal{N}(z^s, (1+z^s)^2 \sigma_z^2)$, imitating a photometric estimate, and uses as prior a physically motivated and simple to calculate gamma distribution with rate parameter $\beta$. In this example, I fix $\beta = 3$, $\sigma_\mu = 0.14$, $\sigma_z = 0.04$ and assume radiation-less $\Lambda$CDM, since the effect of $\Omega_\text{r}$ is negligible at low redshifts. This leaves $\Omega_\text{m0}$ and $\Omega_{\Lambda 0}$ as the free parameters in addition to the SNæ's latent redshifts $\{z^s\}$. The model is depicted graphically in figure 4b. I analyse mock datasets with number of SNæ Ia ranging from 1000 to $10^6$. One dataset is depicted in figure 4a.

---

[23]Cosmological analyses using HMC [33, 53, 65] have all used the `Stan` programming language [14] and its automatic-differentiation capabilities [13], resolving to numerically evaluating the necessary gradients using the Leibnitz integrand rule, applied to eq. (2.3).

[24]It can also be avoided entirely through the use of simulation-based inference (SBI).[25] This family of methods relies on simulating large numbers of mock observations with different underlying cosmological parameters and therefore can benefit from the trivial parallelisation (batching) across various redshifts (for the supernovæ in the dataset) and cosmological parameters (for the different training examples) of the distance routines in `phytorch.cosmology`, which was recently used by [42] in this context. Furthermore, even though calculating the likelihood is not required for SBI, its gradient (if available) can be used to form an addition to the loss function that speeds up training [5, 79] or to derive optimal summary statistics [15].

[25]For overviews of the plethora of methods falling under the umbrella of SBI, see [19, 50].

[26]I use a hat: $\hat{\cdot}$, to denote an observed, measured, or otherwise affected by observational uncertainty quantity, in contrast to the intrinsic (latent) values.

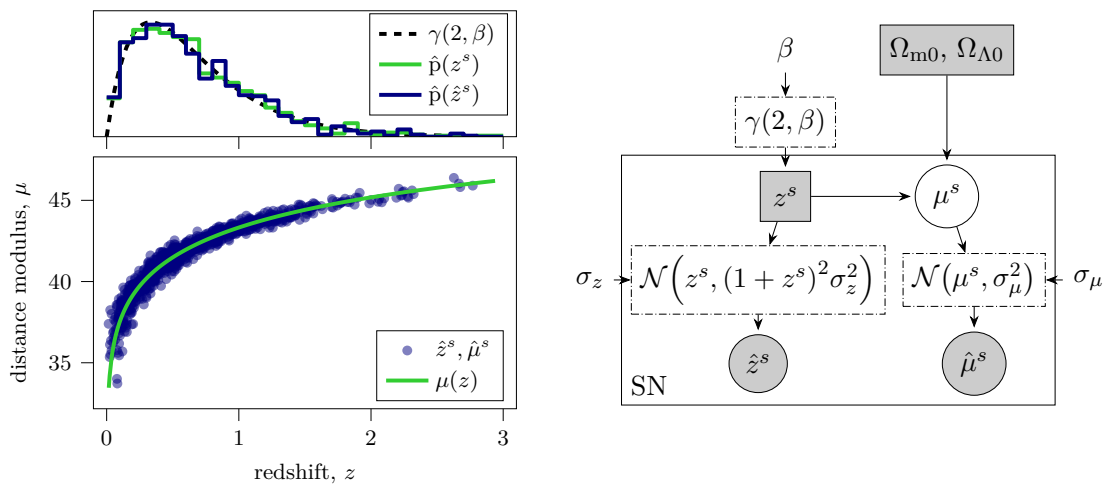

**(a)** Example mock data with 1000 SNæ Ia. *Top*: histograms of the latent and observed redshifts, $z^s$ and $\hat{z}^s$, in comparison with the $z$ prior, also used to draw $z^s$ in the simulator. *Bottom*: observed Hubble diagram as dots and the underlying true relation as a line.

**(b)** Graphical representation of the SN Ia inference model, also used to generate mock data. Shaded boxes indicate parameters, and shaded circles the observed data. $\beta$, $\sigma_\mu$, $\sigma_z$ are the model inputs (fixed parameters) in this work. The plate labelled "SN" indicates the conditional independence of each supernova (with index label $s$).

**Figure 4.** Mock SN Ia data and the model used to generate and analyse it.

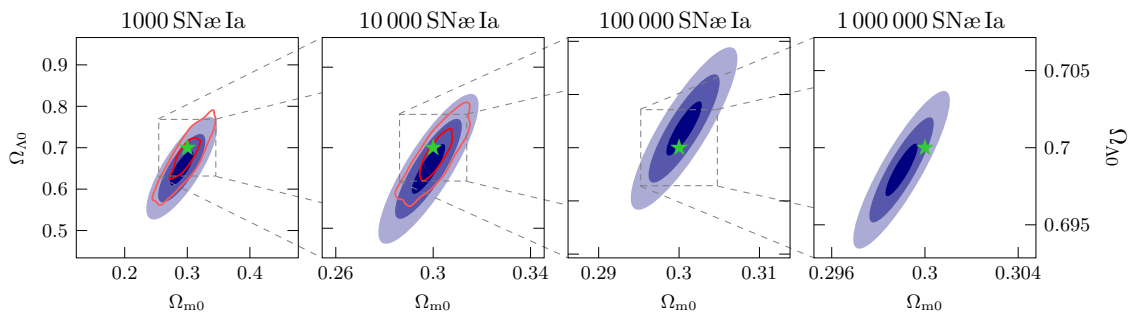

**Figure 5.** Posterior 1-, 2-, and 3-sigma credible regions (with 39.3%, 86.5%, and 98.9% credibility, respectively) for $\Omega_{m0}$ and $\Omega_{\Lambda 0}$ from mock data with increasing numbers of observed type Ia supernovæ. Blue ellipses are results of a VI fit as described in the text, while the red contours were derived with HMC (only performed up to 10000 SNæ Ia and only up to 2-sigma shown). Note the different scales for each plot. The values used to produce the mock data are indicated with a star.

Results from HMC sampling and VI fits to the mock datasets are presented in figure 5. Because of the computational cost of HMC (generating 1000 posterior samples with 10000 SNæ Ia took ≈ 2 h on a high-end workstation), it was only applied to the datasets with 1000 and 10000 SNæ Ia. In contrast, VI can analyse quickly (in ≈ 1 h on the same workstation with an NVIDIA A-100 GPU) up to $10^6$ SNæ Ia. To enable this, I use a *partial multivariate normal* (PMVN) proposal distribution, which accounts for correlations among the two cosmological (global) parameters and between them and each individual SN's latent redshift but ignores additional posterior correlations between different SNæ: full details are given in ([41], subsection 4.2). Due to the conditional structure of the model, a PMVN is sufficient in this case, especially for large observed samples when the cosmological posteriors do approach Gaussianity.

While the particular results of figure 5 are not a focus of this work, we note that VI is successful and efficient for this simple model, with the posterior size shrinking in each dimension as $1/\sqrt{N}$, as expected, and covering the parameter values used to produce the mock data. To the best of my knowledge, this is the first application of VI to cosmological inference with standard candles, albeit with a toy model and mock data. Extending the inference to more realistic models and real datasets, however, requires significant improvements to the guide so that correlations in high dimensions are properly accounted for.[27]

## 5 Conclusion

I have presented analytic expressions for distance calculations in a general (possibly curved) $\Lambda$CDM+radiation cosmology, which utilise the Carlson elliptic integrals, and so can be evaluated using fast numerical algorithms with guaranteed precision. Furthermore, differentiating the Carlson integrals analytically does not require any additional functions, which makes them easy to include in high-performance libraries with automatic differentiation. The formulæ have been implemented in the `phytorch.cosmology` package, part of `phytorch`, which provides `PyTorch`-interfaced GPU kernels for various special functions. Their speed and accuracy have been briefly examined in comparison with numerical integration and other known analytic expressions (applicable only in certain special cases like zero curvature and/or no radiation). Finally, I have demonstrated an example application to cosmological analysis of up to $10^6$ mock type Ia supernovæ using high-dimensional variational inference, which relies on automatically computing gradients through the distance calculations.

## Acknowledgments

I would like to thank Roberto Trotta for encouragement and guidance, John Peacock, Anthony Lasenby, and the anonymous referee for useful comments.

## A Differentiable root-finding

In order to propagate gradients to the cosmological parameters ($\Omega_{\rm r0}$, $\Omega_{\rm m0}$, $\Omega_{\Lambda 0}$), we must differentiate the root-finding operation. Even though both the algebraic solution and eigendecomposition of the companion matrix are automatically differentiable, `phytorch.cosmology` implements explicitly a faster technique. Consider the factorisation of a general polynomial of degree $m$:

$$P(z) = \sum_{k=0}^{m} \alpha_k z^k = \alpha_m \left( z^m + \sum_{k=0}^{m-1} c_k z^k \right) = \alpha_m \prod_{k=1}^{m} (z - r_k), \qquad (A.1)$$

where $c_k$ are the coefficients of the respective polynomial with unit leading coefficient, i.e. $c_k \equiv \alpha_k/\alpha_m$. After differentiating both sides with respect to $c_i$ (note that $i \neq m$ since $c_m = 1$

---

[27]One might consider, for example, using a dense covariance matrix or a normalising flow-based guide, or even train a score-based generative model.

identically, so we cannot differentiate with respect to it) at fixed (or, equivalently, *every*) $z$:

$$z^i = -\sum_{j=1}^{m} \frac{\partial r_j}{\partial c_i} \prod_{\substack{k=1 \\ k \neq j}}^{m} (z - r_k), \tag{A.2}$$

we notice that each product on the right-hand side can be expanded as a polynomial of degree $m-1$, formed by the roots of the original polynomial, excluding $r_j$, i.e. $[r_{i \neq j}]_{i=1}^{m}$, via Vieta's formulæ ([25, 71]; see [74]) (inverse of root-finding):

$$z^i = -\sum_{j=1}^{m} \frac{\partial r_j}{\partial c_i} \sum_{k=0}^{m-1} C_{jk} z^k. \tag{A.3}$$

For each of the $m$ derivatives $\partial/\partial c_i$, eq. (A.3), being of degree $m-1$ and valid for all $z$, implies $m$ equations (one for each coefficient) in the $m \times m$ unknown derivatives $\partial r_j/\partial c_i$. Altogether, they can be summarised in matrix form:

$$\mathbb{1} = \mathbf{C}^\top \frac{\partial \mathbf{r}}{\partial \mathbf{c}}, \tag{A.4}$$

where $\mathbf{r} \equiv [r_i]_{i=1}^{m}$ is to be treated as a column vector, $\mathbf{c} \equiv [c_j]_{j=0}^{m-1}$ as a row vector, and $\mathbb{1}$ is the identity. Thus, we see that the matrix $\mathbf{C}^\top$, formed by the coefficients of the polynomials with roots $[r_{i \neq j}]_{i=1}^{m}$, is in fact the Jacobian of Vieta's formulæ for the original polynomial, i.e. $\mathbf{c}(\mathbf{r})$. By inverting it, we obtain the derivatives necessary for back-propagating through root-finding.

## B More explicit formulæ

### B.1 Cubic case (no radiation)

In the case of vanishing radiation density, $\Omega_{r0} = 0$, the expressions eqs. (3.7) to (3.9) are modified by replacing $\Omega_{r0} \to \Omega_{m0}$ as the leading coefficient and treating one of the linear factors, e.g. $z + a_l \to 1$ (equivalently, changing the label $\square_l$ to $\square_0$). This results, after some simplification possible only in the cubic case, in

$$\chi(z_1, z_2) \xrightarrow{\Omega_{r0}=0} \frac{c}{H_0} \Omega_{m0}^{-\frac{1}{2}} \times 2\Delta z R_f(u_1^2, u_2^2, u_3^2), \tag{B.1}$$

$$t(z_1, z_2) \xrightarrow{\Omega_{r0}=0} \frac{1}{H_0} \Omega_{m0}^{-\frac{1}{2}} \times 2\Delta z \left[ \frac{(\Delta z)^2}{3} R_J(u_1^2, u_2^2, u_3^2, u_{i5}^2) + R_C(s_5^2, q_{i5}^2) \right], \tag{B.2}$$

$$d_{\text{abs}}(z_1, z_2) \xrightarrow{\Omega_{r0}=0} \frac{c}{H_0} \Omega_{m0}^{-\frac{1}{2}} \times \frac{2\Delta z}{3} \Bigg\{ \frac{[\sqrt{z-r_1}\sqrt{z-r_2}\sqrt{z-r_3}]_{z_1}^{z_2}}{\Delta z}$$
$$+ \left[ 3(r_i+1)^2 - d_{ij}d_{ik} \right] R_f(u_1^2, u_2^2, u_3^2)$$
$$+ 2\left( \sum_{\mu=1}^{3} r_\mu + 3 \right) \left[ \frac{(\Delta z)^2}{3} d_{ij} d_{ik} R_D(u_j^2, u_k^2, u_i^2) + \frac{\sqrt{z_1-r_i}\sqrt{z_2-r_i}}{u_i} \right] \Bigg\}, \tag{B.3}$$

with $\{i, j, k\}$ any permutation of $\{1, 2, 3\}$ and

$$u_i \equiv \sqrt{z_2 - r_i}\sqrt{z_1 - r_j}\sqrt{z_1 - r_k} + \sqrt{z_1 - r_i}\sqrt{z_2 - r_j}\sqrt{z_2 - r_k},$$
$$u_{i5}^2 \xrightarrow{\Omega_{r0}=0} u_i^2 + (\Delta z)^2 \times (r_i + 1),$$
$$s_5 \equiv \sqrt{z_1 - r_1}\sqrt{z_1 - r_2}\sqrt{z_1 - r_3}(z_2 + 1) + \sqrt{z_2 - r_1}\sqrt{z_2 - r_2}\sqrt{z_2 - r_3}(z_1 + 1),$$
$$q_{i5}^2 \xrightarrow{\Omega_{r0}=0} (z_1 + 1)(z_2 + 1)u_{i5}^2.$$

### B.2 Legendre form

Here I spell out solutions for the comoving distance and time coordinate in Legendre's basis. From ([27], eqs. (3.147:8) and (3.151:8)):

$$\chi(z_1, z_2) = \frac{c}{H_0}\Omega_{r0}^{-\frac{1}{2}} \times 2\frac{\left[\tilde{F}(c(z), m)\right]_{z_2}^{z_1}}{\sqrt{(r_i - r_k)(r_j - r_l)}}, \tag{B.4}$$

$$t(z_1, z_2) = \frac{1}{H_0}\Omega_{r0}^{-\frac{1}{2}} \times 2\frac{\left[(1 + r_i)\tilde{F}(c(z), m) - (r_i - r_j)\tilde{\Pi}(n, c, m)\right]_{z_2}^{z_1}}{\sqrt{(r_i - r_k)(r_j - r_l)} \times (1 + r_i)(1 + r_j)}, \tag{B.5}$$

with

$$c(z) = \frac{z - r_j}{z - r_i}\frac{r_l - r_i}{r_l - r_j}, \qquad m = \frac{r_k - r_j}{r_k - r_i}\frac{r_l - r_i}{r_l - r_j}, \qquad n = \frac{1 + r_j}{1 + r_i}\frac{r_l - r_i}{r_l - r_j}.$$

Here $\tilde{F}(\phi, m)$ and $\tilde{\Pi}(n, \phi, m)$ are Legendre's integrals of the first and third kind,[28] respectively, and as in ([12], section 19), we have defined $c \equiv \csc^2 \phi$ and used it instead of the argument $\phi$, indicating this with a tilde, as in $\tilde{\Box}(\ldots, c, \ldots) \equiv \Box(\ldots, \phi, \ldots)$.

In the cubic (radiation-less) case eqs. (B.4) and (B.5) reduce, by ([27], eqs. (3.131:8) and (3.137:8)), to:

$$\chi(z_1, z_2) \xrightarrow{\Omega_{r0}=0} \frac{c}{H_0}\Omega_{m0}^{-\frac{1}{2}} \times 2\frac{\left[\tilde{F}(c(z), m)\right]_{z_2}^{z_1}}{\sqrt{r_j - r_i}}, \tag{B.6}$$

$$t(z_1, z_2) \xrightarrow{\Omega_{r0}=0} \frac{c}{H_0}\Omega_{m0}^{-\frac{1}{2}} \times 2\frac{\left[\tilde{F}(c(z), m) - \tilde{P}i(n, c(z), m)\right]_{z_2}^{z_1}}{\sqrt{r_j - r_i} \times (1 + r_i)}, \tag{B.7}$$

with

$$c(z) \xrightarrow{\Omega_{r0}=0} \frac{z - r_i}{r_j - r_i}, \qquad m \xrightarrow{\Omega_{r0}=0} \frac{r_k - r_i}{r_j - r_i}, \qquad n \xrightarrow{\Omega_{r0}=0} -\frac{1 + r_i}{r_j - r_i}.$$

The formulæ eqs. (B.4) to (B.7) are obviously not symmetric in the roots $\{r_i\}$ and are applicable only for certain orderings. The procedure of picking the right assignment of $\{i, j, k\}$, which depends also on the particular limits of integration $z_1, z_2$, is described for the cubic case in [23]. Its non-triviality highlights the utility of using Carlson's symmetric formulation. Similar discussion for the full quartic case is beyond the scope of this article.

---

[28]The notation — argument names and their order — follow the convention of ([39], documentation: https://mpmath.org/doc/current/functions/elliptic.html#legendre-elliptic-integrals).

Finally, by applying the relations

$$F(\phi, m) = R_F(c-1, c-m, c), \tag{B.8}$$

$$F(\phi, m) - \Pi(n, \phi, m) = \frac{1}{3} n R_J(c-1, c-m, c, c-n) \tag{B.9}$$

(ref. [12], section 19, eqs. (19.25.5) and (19.25.14)) to bring eqs. (B.4) and (B.5) to the Carlson form and then the addition theorems ([12], section 19, section 19.26(i)) to combine the two limits of integration, one can re-derive[29] eqs. (3.7) and (3.8) (and similarly, eqs. (B.1) and (B.2) from eqs. (B.6) and (B.7) in the cubic case).

---

[29]Allowing oneself to cancel at will without regard for such things as branch cuts.